\documentclass[twocolumn,notitlepage,floatfix]{revtex4-2}
\usepackage[english]{babel}
\usepackage[utf8x]{inputenc}
\usepackage{subcaption}

\usepackage{amsmath}
\usepackage{amssymb}
\usepackage{graphicx}

\begin{document}
\title{Nanopore and tunneling resonance in an open kink in  carbon nanotubes}
\author{Alex Kleiner \\orcid.org/0000-0002-4694-2218}
\date{March 10, 2023}

\begin{abstract}
In  weakly buckled carbon nanotubes,  the kink is an open, ovalized  constriction -- where molecular-sized analytes and ions can pass.  Moreover, in  semiconducting tubes, the kink has an electronic bound state within the bandgap. This state is localized, hence  highly susceptible to electrostatic perturbation by a proximate, itinerant ion. The degree to which this perturbation will modulate the electronic current -- which is our signal, can be maximized in a tunnel-junction set-up, near resonance.
We thus propose such a device, analyse its performance  and, since the bound state depends on the post-buckling bending, suggest a scalable fabrication method to achieve bending and buckling of carbon nanotubes to an adjustable degree. 
\end{abstract}

\maketitle

\subsubsection{Introduction}
Solid-state nanopore  devices may revolutionize medicine \cite{Dekker_review_2022} but have poor signal-to-noise ratios compared with biological ones \cite{dekker_noise}; their pores, fabricated by top-down methods, are inherently larger and rougher. In addition, all label-free nanopore devices work by ionic-current blockade, which is hard to scale.

Here we propose a scalable solid-state nanopore device with a smooth and adjustable "pore" where the signal is electronic. The proposed pore is a
 shallow, ovalized kink (hereafter: open kink), in a buckled semiconducting carbon nanotube (CNT). The highly localized  kink deformation affects both ionic and electronic currents: 
depending on the tube type and degree of curvature, it may form a potential well or barrier \cite{my_paper_fundamental} for conduction electrons.   
Hence, a passing ion, by effectively gating the surrounding local potential, can modify the electronic current, especially when the current is due to tunneling.  This concept can be applied, in principle, to all semiconducting tubes (baring in mind the differences \cite{my_paper_fundamental}), but  throughout this work, the derived results are quantitatively exemplified with  a specific  tube: the zigzag (10,0). 

First, however, we describe the proposed method of bending and buckling the tube so that its final bending curvature (and thus, the depth of the kink's localized bound state) could be determined in the initial lithographic patterning stage. 


\begin{figure*}[]  
\begin{center}
\includegraphics[width=0.85\textwidth]{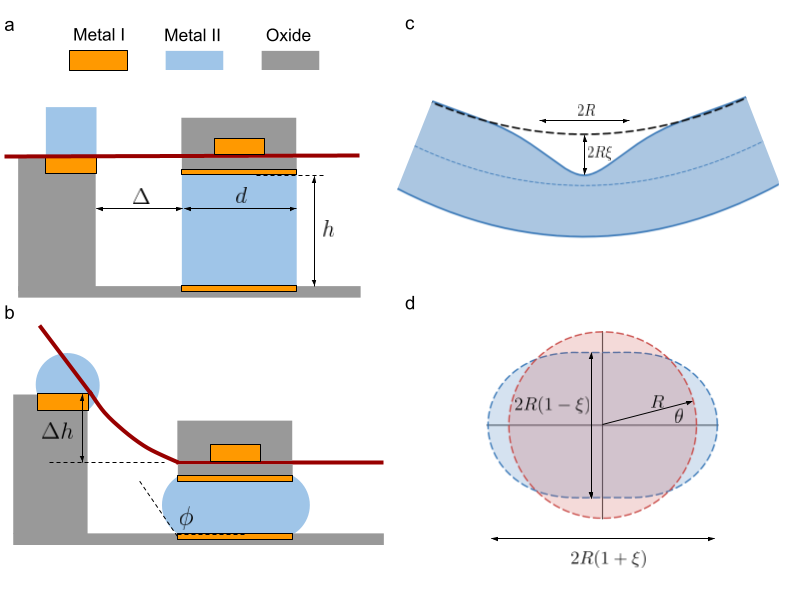}\caption{(a)-(b) Main two steps. Metal II has a low melting-point $T_m$, but the structure was deposited at $T<T_m$. On raising the temperature $T>T_m$, metal II melts and wets metal I (which remains solid and fixed to the oxide substrate). Capillary forces  pull down the structural layer including the tube. (c) Open kink profile. The kink's width $W\sim 2R$. (d) Kink's cross-section as parametrized by the ovalization parameter $\xi$, having    $R\rightarrow R(1+\xi\cos{2\theta})$. At the onset of buckling its already slightly ovalized at $\xi=2/9$ (blue shape). The closing  of the kink corresponds to $\xi^\textrm{close}$ (eq. \ref{eq:xi_close}).
} \label{fig:layout} 
\end{center}
\end{figure*}

\subsubsection{Layout of the device}
In the following we discuss  the physical  realization of the kink junction (fig. \ref{fig:layout}, see also  \cite{my_patent}). Consider a tube deposited on a pre-patterned trench or step as in fig. (\ref{fig:layout}a). 
In it, there are two metal types: metal I, which is a "normal" metal with high melting point, and metal II with a low melting point (but above room temperature), such as Gallium. The tube is then  clamped to the column in the trench with a normal metal electrode (metal I) and additional layer of oxide (together, forming the so-called structural layer). The entire layout in fig. (\ref{fig:layout}a) is assumed to be grown  at a lower temperature then the melting point of metal II.

Upon raising the temperature above the melting point of metal II, the droplet adopts a new conformation (fig. \ref{fig:layout}b) with the wetting angle $\phi$ given by,
\begin{equation}
\gamma_{II}\cos{\phi}+\gamma_{I-II}=\gamma_o,
\end{equation}
where $\gamma_{II},\gamma_{I-II},\gamma_o$ are the surface tensions of metal $II$, interface of metals I and II and oxide, respectively.

\begin{figure}[]  
\begin{center}
\includegraphics[width=0.5\textwidth]{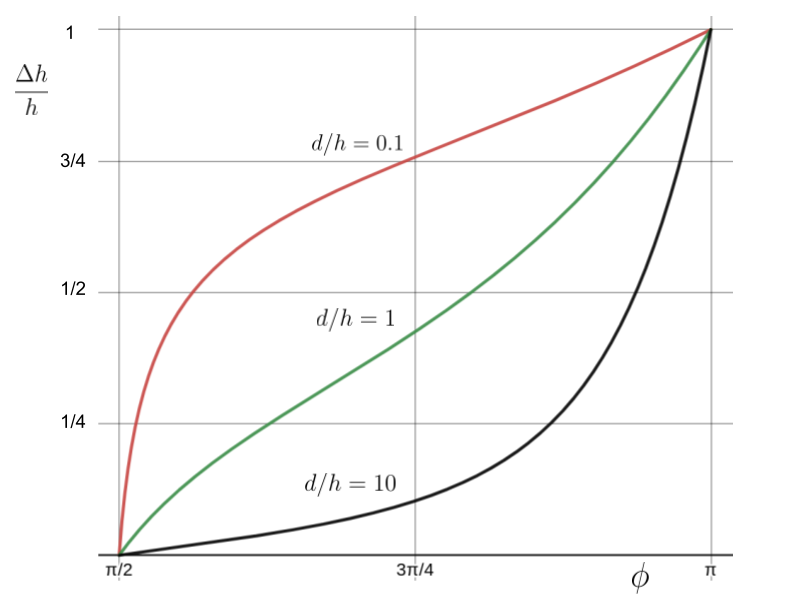}\caption{Change of height vs. wetting angle $\phi$ for a number initial aspect ratios. 
} \label{fig:delta_h} 
\end{center}
\end{figure}

Assuming the in-plane dimension of the layout is $\gg d$ and large enough so that resistance from the bending tube does not notably affect the final conformation, then by elementary geometry we get,
\begin{equation}\label{eq:delta_h}
    \Delta h=h+\frac{1}{2}\left( f(\phi)d-\sqrt{4hf(\phi)d+f(\phi)^2d^2}\right),
\end{equation}
where 
\begin{equation*}
    f(\phi)=\frac{4\sin^2\phi}{2\phi-\pi-\sin{2\phi}}.
\end{equation*}
Fig. (\ref{fig:delta_h}) depicts the height change for a number of aspect ratios according to eq. (\ref{eq:delta_h}).

The bent section (fig. \ref{fig:layout}b), at the onset of buckling, has a curvature of
\begin{equation}
    \kappa=\frac{2\Delta h}{(\Delta h)^2+\Delta^2}.
\end{equation}
The critical curvature at which CNTs buckle was found by MD simulations to depend on the tube radius in accordance with  continuous elasticity theory, 
\begin{equation}\label{eq:kappa_cr}
    \kappa^\textrm{cr}=\frac{A}{R^2}
\end{equation}
where the value of $A$ found by different groups somewhat varies ($A=0.385$\AA\, by \cite{yakobson1},  $0.376$\AA\, by \cite{poisson2002} and  $0.185$\AA\, according to \cite{cao}; see also \cite{like_yakobson}). 
Thus, in order to induce buckling,  ($\kappa\ge\kappa^\textrm{cr}$), the height of the structural layer $h$ (fig. \ref{fig:layout}b) must be lowered  by,
\begin{equation}\label{eq:h}
    \Delta h\ge\frac{R^2}{A}-\sqrt{\frac{R^4}{A^2}-\Delta^2},
\end{equation}
where $\Delta \leq R^2/A$.  To see what $h$ and $\Delta$ might be, consider tubes in the range  (10,0) to (20,0), having $R=4$\AA\, and $8$\AA, respectively. Taking $A = 0.185$\AA\,\cite{cao},  eq. (\ref{eq:h}) gives $h=\Delta$, where   $\Delta\leq 8.65$nm for the (10,0) tube, and   $\Delta\leq 34.6$nm, for the (20,0) one. These are rather small values but we need to keep in mind that  local imperfections are likely to reduce the critical curvature, hence the values given for $A$ here should be treated as an ideal \emph{upper} limit.

\subsubsection{Open kink}
The elasticity of carbon nanotubes was found, both experimentally \cite{ijima}\cite{treacy} and by simulations \cite{ijima}\cite{yakobson1}\cite{like_yakobson}\cite{cao}\cite{poisson2002}\cite{zhang2006}, to conform to the elasticity of  thin cylindrical shells within continuous elastic theory  \cite{shell_book}.
Their properties under bending and buckling were explained by Brazier  a century ago \cite{brazier}. In a previous work  we summarize this theory as applied to single-walled CNTs (\cite{my_paper_fundamental} - appendix A).

MD-simulations of bent CNTs \cite{transient_deformation} uncovered two critical points: the buckling curvature 
 $\kappa^\textrm{cr}$ which marks the onset of the kink, and a second curvature $\kappa^\textrm{close}$, which marks the closing of the  kink's cross-section. 
 The region between them is called the \emph{transient regime} \cite{transient_deformation}. It is characterized by having an open kink  with an ovalized, bending-dependent,  cross-section.
 This work requires an open kink and thus lies entirely within the transient regime.
 
 The ovalization parameter of the open kink, $\xi^\textrm{kink}$, is given by \cite{my_paper_fundamental},
\begin{equation}
\label{eq:xi_kink}
    \xi^\textrm{kink}=\frac{2}{9}+\left(\xi^\textrm{close}-\frac{2}{9}\right)\tilde{\kappa}^{1/2},
\end{equation}    
    where
    \begin{equation}\label{eq:kappa_tilde_def}
    \tilde{\kappa}\equiv \frac{\kappa-\kappa^\textrm{cr}}{\kappa^\textrm{close}-\kappa^\textrm{cr}},\,\, \textrm{having}\,\,\kappa^\textrm{cr}\le\kappa\le\kappa^\textrm{close}, 
\end{equation}
and $\kappa^\textrm{close}$ is  the bending curvature at which
the kink closes: i.e: its ovalization parameter,
\begin{equation}\label{eq:xi_close}
    \xi^\textrm{close}\equiv 1-\frac{d_g}{2R},
\end{equation}
 corresponds to maximum flattening of the cross-section: i.e: where the distance between the opposite walls of the kink equals the inter-layer distance in graphite ($d_g=3.35$\AA). A (10,0) tube, with $R=4$\AA, has a $\xi^\textrm{close}\approx 0.58$.   Since at the onset of buckling $\xi=2/9$ \cite{brazier},  the transient regime of the kink, which is the focus of this work, is thus given by
 \begin{equation}\label{eq:transient_xi}
     \frac{2}{9}<\xi^\textrm{kink}<\xi^\textrm{close},
 \end{equation}
 or equivalently, by (\ref{eq:kappa_tilde_def}),
 \begin{equation}\label{eq:kappa_tilde}
     0<\tilde{\kappa}<1.
 \end{equation}

\subsubsection{Kink's potential}

The electronic band-structure of bent and buckled tubes was studied by many authors and analysed at depth recently by us (see \cite{my_paper_fundamental} and citations therein).  

To summarize it briefly, pure bending has no net strain but it does affect the bandgap as follows: bending causes ovalization of the cross-section which increases the overall circumferential curvature -- which is what affects the bandgaps;  it was shown \cite{my_paper_fundamental} that bandgaps, initially $\propto \cos{3\alpha}\,/R^2$, become $\propto (1+\frac{9}{2}\xi^2)\cos{3\alpha} / R^2$, where $\xi$ is the ovalization parameter (fig. \ref{fig:layout}d). In the pre-buckling regime, $\xi\propto\kappa^2$ so that the change due to bending is $\propto\kappa^4$. At post-buckling, however, the bulk (not including the kink) is independent of further bending and remains at the  critical ovalization, $\xi=2/9$; while the at kink $\xi^\textrm{kink}-2/9\propto\tilde{\kappa}^{1/2}$ (eq. \ref{eq:xi_kink}). Hence, not too far from criticality, the bandgap at the kink changes as $\propto (\xi^\textrm{kink}-2/9)^2 \propto\tilde{\kappa}^{1/2}\cos{3\alpha}\,/R^2$.

\begin{figure}[]  
\begin{center}
\includegraphics[width=11cm]{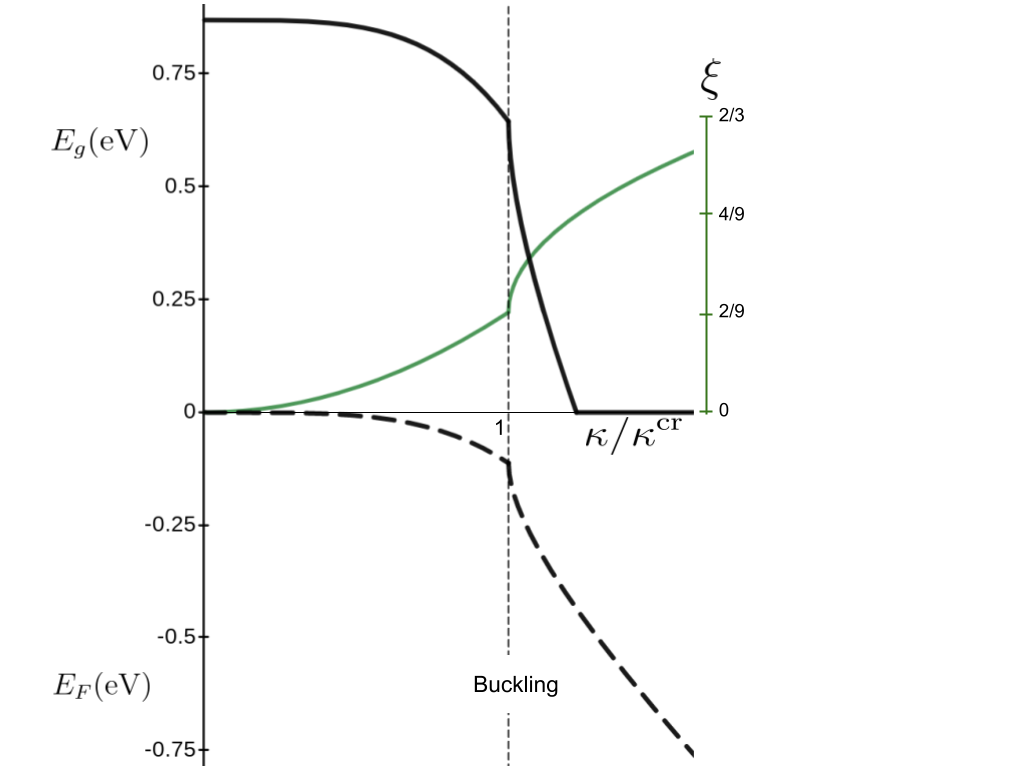}\caption{The bandgap $E_g$ (full line), Fermi energy $E_F$ (broken line) and ovalization parameter $\xi$ (green) of a (10,0) tube under bending $\kappa$. The post-buckling section $(\kappa / \kappa^\textrm{cr} > 1)$ corresponds to the kink.   
} \label{fig:10_0} 
\end{center}
\end{figure}

With the above in mind, we  apply next the general analysis in \cite{my_paper_fundamental} to   the zigzag (10,0). Its bandgap, Fermi energy and ovalization as a function of bending curvature are given in fig. (\ref{fig:10_0}). For tubes of this size ($R=4$\AA) and smaller, Fermi energy downshifts in tandem with the bandgap as $\Delta E_F=\Delta E_g / 2$, which is $\propto\kappa^4$ at the pre-buckling stage; its physical origin is the increased $\sigma-\pi$ hybridization at large circumferential curvatures, which at sufficiently large bending vanishes the bandgap entirely and downshifts the Fermi energy.  

Fig. (\ref{fig:10_0}) also reveals the evolution of the ovalization parameter $\xi$; at pre-buckling, $\xi\propto\kappa^2$ up to the onset of buckling -- where $\xi=2/9$ \cite{brazier}; at post-buckling, away from the kink $\xi$ remains constant  while at the kink, $\xi^\textrm{kink}\propto\tilde{\kappa}^{1/2}$, where $\tilde{\kappa}$ is the post-buckling dimensionless bending (eq. \ref{eq:kappa_tilde}).
One can also observe that the bandgap at the kink vanishes when the its ovalization is $\xi^\textrm{kink}\gtrapprox 4/9$. That will

 For the (10,0) tube (and other tubes of similar radius), Fermi energy was found to downshift due to bending-induced ovalization by
\begin{equation}\label{eq:EF_xi_final}
     E_F(\xi)=\frac{9C_s\xi^2}{4R^2}\cos{3\alpha},\,\,\,\,\,
 \end{equation}
 where  $C_s=8$ (\AA$^2\cdot $eV), $\alpha$ is the chiral angle and 
 $\xi$ is the ovalization parameter (fig. \ref{fig:layout}d). In buckled tubes 
 the kink is highly ovalized results in a difference in Fermi energy with the bulk,
 \begin{equation}\label{eq:EF_diff}
     \Delta E_F=\frac{9C_s}{4R^2}\left(\left(\frac{2}{9}\right)^2-(\xi^\textrm{kink})^2\right)\cos{3\alpha},
 \end{equation}
 where the ovalization parameter at the bulk is, at the post-buckling stage, is fixed roughly at its value at the onset of buckling $\xi=2/9$ \cite{brazier}.

The potential difference between the kink and the bulk is thus,

\begin{equation}\label{eq:Vk_def}
    V_\kappa=-\Delta E_F+\frac{1}{2}\Delta E_g=-2\Delta E_F,
\end{equation}

Substituting eq. (\ref{eq:xi_kink}) in (\ref{eq:EF_diff}), eq. (\ref{eq:Vk_def}) becomes
\begin{eqnarray}\label{eq:Vk}
    V_\kappa&=&-\frac{2C_s\left(\xi^\textrm{close}-\frac{2}{9}\right)\tilde{\kappa}^{1/2}}{R^2}\cos{3\alpha},\nonumber\\
    &\approx&-\frac{2C_s\tilde{\kappa}^{1/2}}{3R^2}\cos{3\alpha},
\end{eqnarray}
where in the second row we substituted $\xi^\textrm{close}$ for its value for the (10,0) tube (see text after eq. \ref{eq:xi_close}).

\begin{figure}[]  
\begin{center}
\includegraphics[width=11cm]{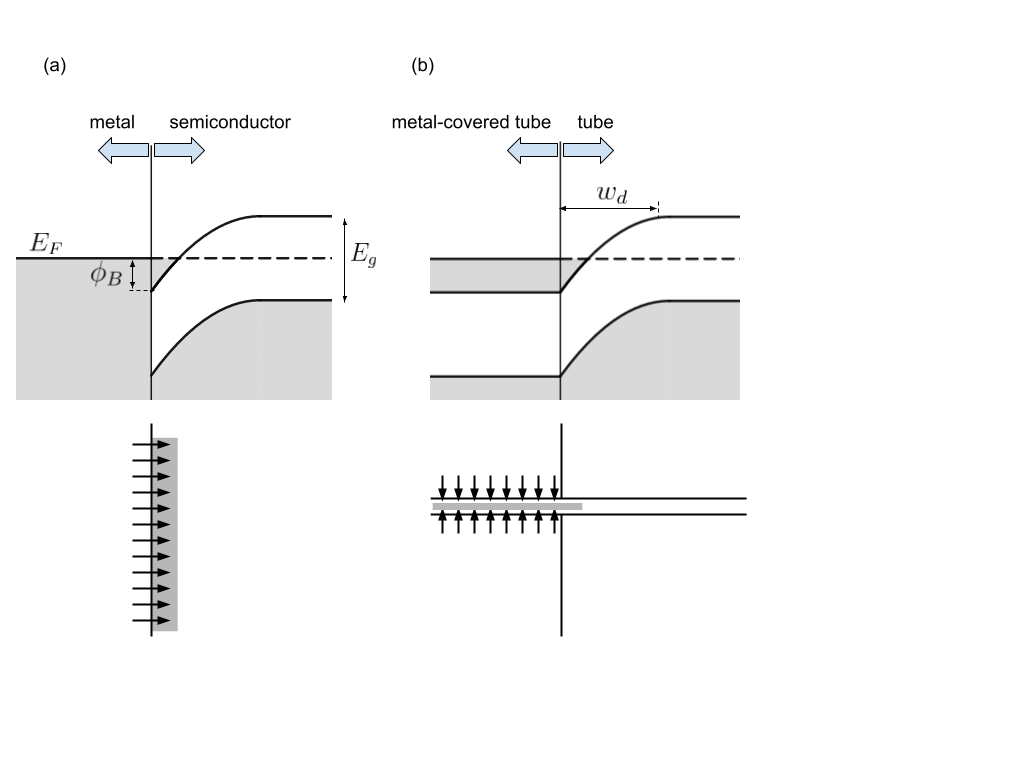}\caption{Comparing conventional metal-semiconductor interface (a) with a one with semiconducting CNT (b). Lower panels show the respective orientation of the interface dipoles. 
} \label{fig:contact} 
\end{center}
\end{figure}

\subsubsection{Bound state}

The width $W$ of the kink potential $V_k$ (eq. \ref{eq:Vk}) is of the order of the tube diameter $\sim$1nm. If the Fermi wavelength is much larger, the potential can be approximated to a  $\delta$-function. Indeed, a class of tubes have their $k_F$ zone-folded to zero: tubes that satisfy \cite{Jishi1994}

\begin{equation}\label{eq:long_gcd}
    \textrm{gcd}\left(2m+n, 2n+m\right) =\textrm{gcd}(n,m),
\end{equation}
where \textrm{gcd} is the greatest common divisor. All zigzag tubes ($m=0$), for example, meet this condition. 
Hence, for these tubes, the actual wavelength of the tunneling electrons is the thermal length, $\lambda_\textrm{th}=\sqrt{2\pi\hbar^2/m^*kT}$, where
 $m^*$ is the effective mass at the band-edge:   $m^*=\hbar^2(\partial^2E/\partial k^2)^{-1}= 2\hbar^2\Delta k_y/(a\gamma\sqrt{3})$, where the energy spectrum of the $\pi$ band, $E=\frac{\sqrt{3}}{2}a\gamma\sqrt{\Delta k_y^2+k^2}$ was used (for simplicity, the effective mass is assumed not to change significantly by mixing with the singlet band); now since $\Delta k_y=E_g/(a\gamma\sqrt{3})$, one gets,
\begin{equation}\label{eq:mass}
    m^*=\frac{2\hbar^2E_g}{3a^2\gamma^2},
\end{equation}
which gives, for $E_g=1$eV, $m^*=0.09m_e$, where $m_e$ is the electron mass; the thermal length is then
\begin{equation}\label{eq:labda_thermal}
    \lambda_\textrm{th}=\sqrt{\frac{3\pi a^2\gamma^2}{E_g kT}},
\end{equation}
giving $\lambda_\textrm{th}\approx 29$ nm at liquid-nitrogen temperatures while  $\lambda_\textrm{th}\approx 15$ nm at room-temperature.
Hence, even at room temperature $\lambda_{th}\gg W$, thus justifying its $\delta$-function approximation.

Thus, considering the kink potential well as a $\delta$-function with height $V_\kappa$ (eq. \ref{eq:Vk}) and width $W=2R$, 
the  wavefunction of the bound-state is then given by
\begin{eqnarray}\label{eq:psi}
    \psi&=&\sqrt{\frac{c}{2\pi R}}\exp{\left(-c|z|\right)},\,\,\,\,\,\,\textrm{where}\nonumber\\
    c&=&\frac{4E_gV_\kappa R}{3a^2\gamma^2}.
\end{eqnarray}
A (10,0) tube, for example, having $E_g\approx 1$eV, $V_\kappa=2\Delta E_F\approx 1$eV (see fig. \ref{fig:10_0})  and $R=4$\AA,  gives $c\approx 1$nm$^{-1}$. 

The $\delta$-function potential of area $V_\kappa W$ has at least one bound state of energy
\begin{equation}\label{eq:E01}
    E_\kappa=-\frac{m^*V_\kappa^2W^2}{2\hbar^2}=-\frac{4E_gV_\kappa^2R^2}{3a^2\gamma^2}
\end{equation}
where the effective mass was substituted by eq. \ref{eq:mass}. For the same (10,0) tube and bending as in the example after eq. (\ref{eq:psi}), this gives $E_\kappa=-0.38$eV.  


\subsubsection{Single ion perturbation}
How is an ion passing through the open kink will affect the current? The surface potential due to an ion  of charge $-1e$ at the center of the kink is
\begin{equation}
    \Phi=\frac{k_ee^2}{\epsilon_r\sqrt{R^2+z^2}}
\end{equation}
where $k_e=8.987\times 10^9\textrm{N}\cdot$m/C$^2$, and 
$\epsilon_r$ is the relative dielectric permittivity.  It perturbs  the bound state  by
\begin{eqnarray}\label{eq:Delta_E0}
    \Delta E_\kappa&=&\int \psi^2\Phi dz\nonumber\\
    &=&2\pi R W \psi^2(0)\Phi(0)=\frac{8k_ee^2E_gV_\kappa R}{3\epsilon_ra^2\gamma^2}
\end{eqnarray}
where in the second row $\Phi$ was taken as a local potential (as we did with $V_\kappa$) extending throughout a ring of width $W$ at the center of the kink.
Returning to the (10,0) tube and applying eq. (\ref{eq:Delta_E0}), we get  $\Delta E_\kappa\sim 0.27$\,eV for a typical value of $\epsilon_r\approx 10$. This is  a large perturbation as it is  comparable with the zero-order energy level $E_\kappa=0.38$eV  (after eq. 
\ref{eq:E01}).

\subsubsection{Contacts}
The electronic transport depends, to a large extent, on the metal-nanotube contact. Experiments with such contacts demonstrated the differences with conventional metal-semiconductor interface but also resulted often in contradictory conclusions \cite{Campbell2011}. Some of this could be attributed to the fact that  while tube diameter was usually known, its chiral angle was not (semiconducting tubes of equal diameter but different chirality  give different work-functions \cite{my_paper_fundamental}).  

Nevertheless,  the nature of the metal-CNT contact can be outlined and compared with the  much-studied  bulk metal-semiconductor interface (fig. \ref{fig:contact}a). Here, different work-functions across the interface result in Schottky barriers \cite{sze} 
\begin{eqnarray}\label{eq:phi_B}
    \phi_{B,n}&=&\phi_m - \chi -eD_{\textrm{int}}\\
    \phi_{B,p}&=&E_g-\phi_{B,n}
\end{eqnarray}
where $\phi_{B,n}$ or $\phi_{B,p}$ are the Schottky barriers for $n$ or $p$ type semiconductors, $\chi$ is the semiconductor's electron affinity, $\phi_m$ the metal work-function and $D_{\textrm{int}}$ is the density of electrostatic dipoles at the interface -- a consequence of  metallic
 Bloch  surface states that  decay into the semiconductor (in the so-called electron-accumulation region).

 Consider now the metal-CNT contacts, where the  CNT  is covered  by a metal electrode (fig. \ref{fig:contact}b).
 It was studied, for example, by measuring the Schottky barriers at contacts between the same CNT and electrodes of different materials -- Hf, Cr, Ti and Pd  \cite{perello2010}. It was found that electrodes with low work-function such as Hf (where metal-CNT dipoles should be largest), induce a conducting region within the metal-covered nanotube, akin to the electron-accumulation region in bulk interfaces. This  seem to extend within the metal-covered region of the tube up to about 100nm \cite{franklin_length}. Next to it, in the un-covered segment of the nanotube, lies the ideal Schottky barrier, extended throughout the depletion length ($w_d$ in fig. \ref{fig:contact}b). 
 
 Thus, the resulting
 electron transport at the metal-CNT interface consists of crossing two barriers in series: first, tunneling between the metal and the conduction region within the metal-covered tube (which takes place anywhere within a distance of $\sim$ 100nm from the free tube), and second, tunneling between this region and the free tube across the Schottky barrier.

 \subsubsection{Depletion region}
Let us now estimate the depletion region in the free tube near an electrode. Say the electrode covers the tube  at $z\le0$, then Poisson equation gives
\begin{equation}
    -\frac{d^2\phi}{dz^2}=\frac{\rho(z)}{\epsilon}
 \end{equation}   
and the charge density $\rho$ is assumed constant in each of the two segments,
\begin{equation*}
\rho(z)=
        \begin{cases}
            -eN_m,& \,\,\,\,\,\,\,z\le0\\
            eN_f,& \,\,\,\, 0<z<w_d
        \end{cases}    
\end{equation*}
where $N_m$ and $N_f$ are the charge densities in the segment covered by metal and free, respectively, and $w_d$ is the depletion region. Solving for $\phi$ one gets, 
\begin{equation}
    w_d=\left(\frac{2\epsilon \phi_B}{e\left(N_f+\frac{N_f^2}{N_m}\right)}\right)^{1/2}
    \end{equation}
Assuming the tube is not doped, with $E_g\sim 1$eV, $N_f$ is given by
\begin{equation}
    N_f=\int_{E_c}^\infty g(E)e^{-(E-E_c)/kT}dE
\end{equation}
where $g(E)$ is the  density of states and
 $E_c$ is the conduction band energy above the chemical potential. Now since  $E_c-E_F\gg kT$ (in 10,0 tube, $E_c-E_F\sim 0.45$eV $\gg kT$  even at room temperature), $w_d\gtrapprox 10^{-7}$m. This region is much larger then the length where tunneling can be effective. Thus, the tunnel junctions  length $l$ is assumed here to be $l\ll w_d$.

 \begin{figure}[]  
\begin{center}
\includegraphics[height=0.4\textheight]{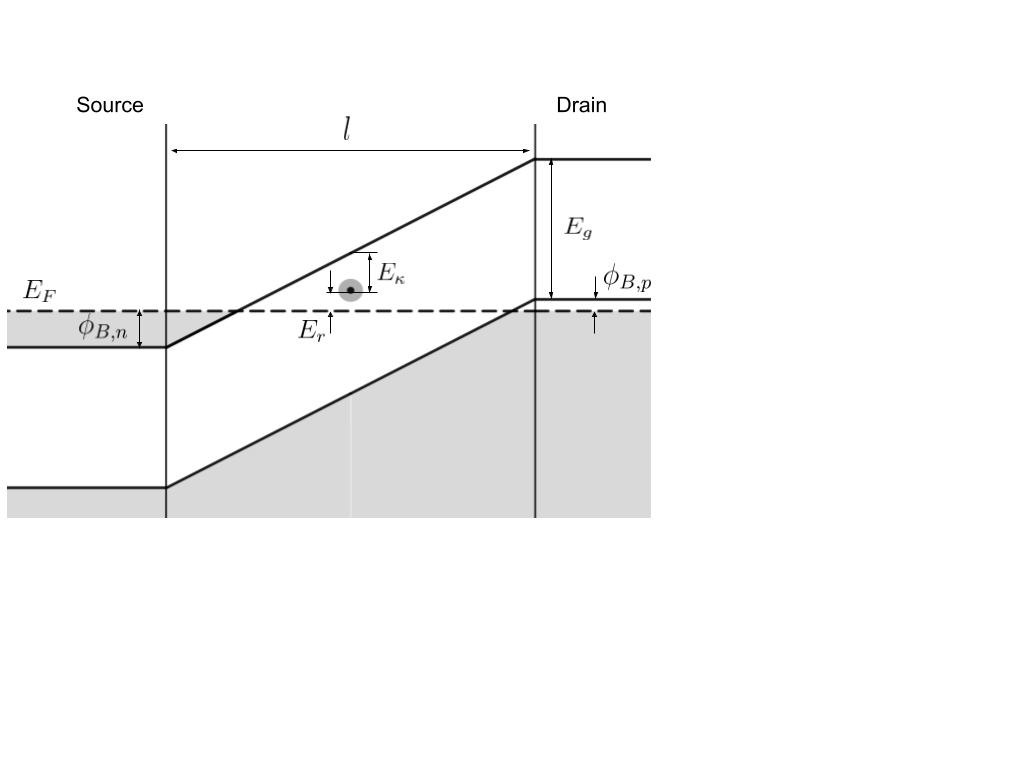}\caption{  A  semiconducting tube in an asymmetric junction: the source is a low work-function metal, such as Hafnium, and the drain is metal with a small barrier, such as Pd.  The depicted junction is short ($l\ll w_d$). 
} \label{fig:thermionic_vs_tunneling} 
\end{center}
\end{figure}

\begin{figure}[h]  
\begin{center}
\includegraphics[width=0.7\textwidth]{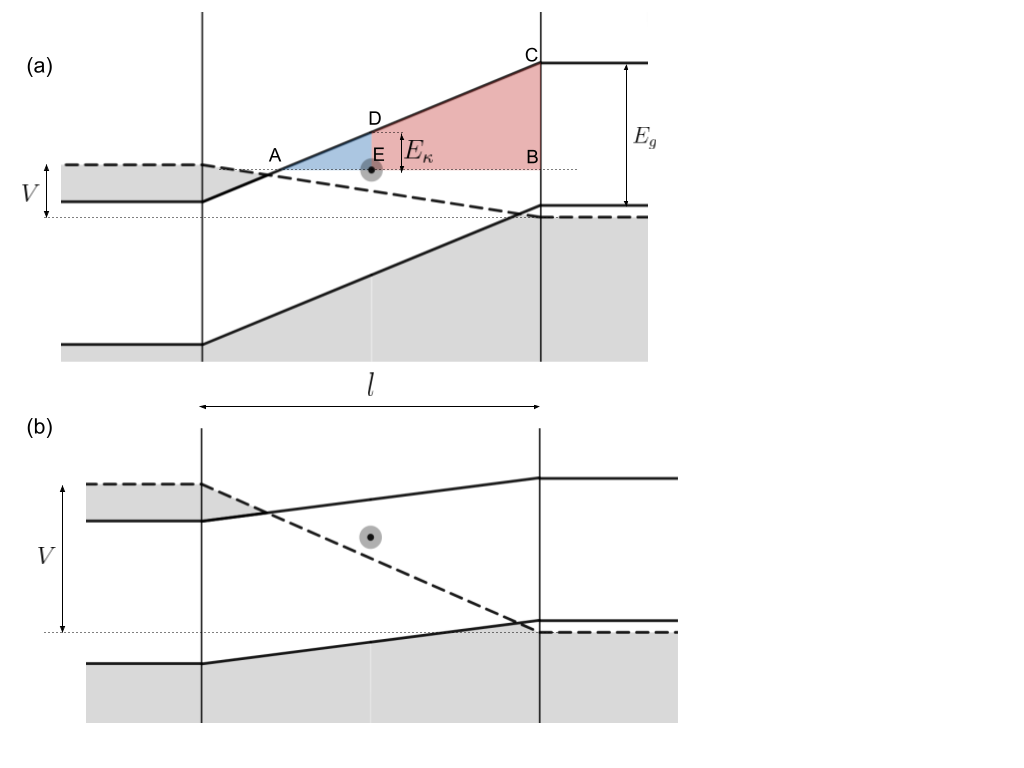}\caption{Band structure vs. drain-source bias at resonance (a) and valley (b). The resonance condition is given by eq. (\ref{eq:resonance_range}). The blue and red areas are tunneling barriers with transmissions given by eqs. (\ref{eq:tunnelings_to_kink}) and (\ref{eq:tunnelings_from_kink}), respectively.} \label{fig:band_vs_bias} 
\end{center}
\end{figure}

 \subsubsection{A short asymmetric junction}
 
Fig. \ref{fig:thermionic_vs_tunneling} depicts an asymmetric short junction ($l\ll w_d$), where the source and drain consists of metals with low and high work-functions, respectively (see appendix for work-functions).

Ignoring for the moment the kink and its bound state, charge transport in this junction (fig. \ref{fig:thermionic_vs_tunneling}) consists of the following. First, tunneling from the metal electrode at the source into the  metal-induced $n$-states in the nanotube (of depth $\phi_{B,p}$), then tunneling through the free section of the tube into the metal-induced $p$-states in the tube, and finally tunneling out to the metal electrode at the drain.    

Now adding the kink -- with its bound state at depth $E_\kappa$ below the conduction band -- and a bias $V$ between source and drain  (fig. \ref{fig:band_vs_bias}), we get the following conduction regimes.   The height of the kink state above the Fermi energy at the source is 
\begin{equation}\label{eq:Er}
    E_r=\frac{1}{2}(E_g+\phi_{B,p}-\phi_{B,n}-V)-E_\kappa
\end{equation}
 while relative to  the conduction edge it is $E_r+\phi_{B,n}$. This sets the range of the "On" state, $-\phi_{B,n}\leq E_r \leq 0$, that is
\begin{equation}\label{eq:resonance_range}
  -\phi_{B,n} \leq 2E_\kappa-E_g-\phi_{B,p}+V\leq \phi_{B,n}
\end{equation}
When, at a larger bias, the junction is  out of the "On" range, the bound state in the kink lies under the conduction band edge (fig. \ref{fig:band_vs_bias}b), and the current decays exponentially. 

A point of special interest is where the  negative differential resistance is maximum. This occurs where $E_r$ crosses the  band edge of the source, which is just beyond the right-hand inequality of the resonance condition (\ref{eq:resonance_range}),
\begin{equation}\label{eq:E0_resonance}
    2E_\kappa\gtrapprox E_g+\phi_{B,p}+\phi_{B,n}-V.
\end{equation}
Eq. (\ref{eq:E0_resonance}) gives the bias region at which the junction's negative differential resistance peaks. Hence, at this point the current's response to the ionic perturbation of the  bound state is highest.

\subsubsection{Current and ionic perturbation}

In our asymmetric junction, the direct tunneling between source and drain, ignoring the kink, would be under the triangle ABC (fig. \ref{fig:band_vs_bias}a). The energy BC is $E_g-E_r$ (eq. \ref{eq:Er}) which is, within the WKB approximation and at zero bias, 

\begin{eqnarray}
    T_{sd}&\approx& \exp{\left(-2\int_\textrm{A}^{\textrm{B}}|k(z)|dz\right)}\nonumber\\
    &=&\exp{\left(-\frac{4\sqrt{2}l}{3\hbar}\sqrt{m^*(E_g-E_r)}\right)}\nonumber\\
  &=&\exp{\left(-\frac{4l(E_g+2E_\kappa-\phi_{B,p}+\phi_{B,n}+V)}{3\sqrt{3}a\gamma}\right)}\nonumber\\
  &\approx&\exp{\left(-\frac{8l(E_g+\phi_{B,n})}{3\sqrt{3}a\gamma}\right)}, 
\end{eqnarray}
where the last step assumes near resonance (eq. \ref{eq:E0_resonance}).

Within the "On" range (eq. \ref{eq:resonance_range}), the tunneling between source and kink (at the center of the junction: fig. \ref{fig:band_vs_bias}a)  is under the triangular potential ADE, which is, 
\begin{eqnarray}\label{eq:tunnelings_to_kink}
    T_{s, k}&\approx& \exp{\left(-2\int_\textrm{A}^{\textrm{E}}|k(z)|dz\right)}\nonumber\\
    &=&\exp{\left(-\frac{4l}{3\sqrt{2}\hbar}\sqrt{m^*E_\kappa}\right)}\nonumber\\
    &=&\exp{\left(-\frac{8lE_gC_s\sqrt{\tilde{\kappa}}\cos{3\alpha}}{9Ra^2\gamma^2}\right)}    
\end{eqnarray}
where in the last step we substituted eq. (\ref{eq:mass}) for $m^*$ and eq. (\ref{eq:E01}) for $E_\kappa$.

\begin{figure}[]  
\begin{center}
\includegraphics[width=0.65\textwidth]{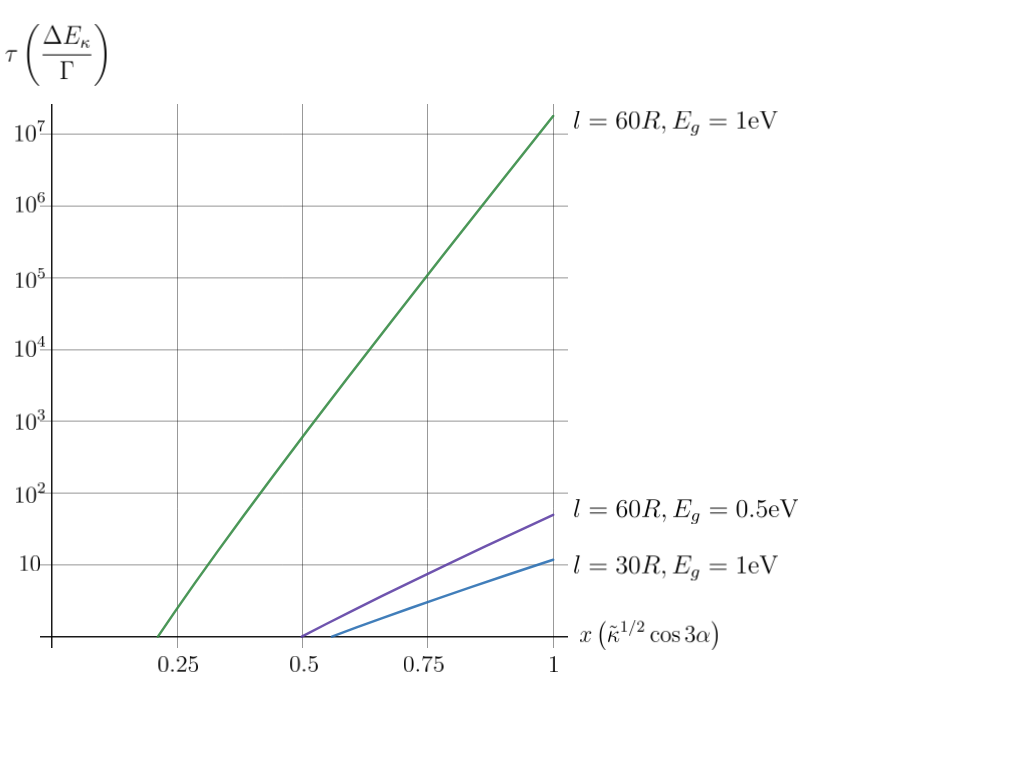}\caption{ $\tau$ is a dimensionless shift of the bound  level due to a single  ion inside the kink, plotted here vs. post-buckling curvature   $x\equiv \tilde{\kappa}^{1/2}\cos{3\alpha}$, for a number of junction lengths $l$ and bandgaps.} \label{fig:tau} 
\end{center}
\end{figure}

The tunneling between kink and drain (red area in fig. \ref{fig:band_vs_bias}a) is 

\begin{eqnarray}\label{eq:tunnelings_from_kink}
    T_{k, d}&\approx& \exp{\left(-2\int_\textrm{E}^{\textrm{B}}|k(z)|dz\right)},\nonumber\\
    &=&\exp{\left(-2\left(\int_\textrm{A}^{\textrm{B}}|k(z)|dz-\int_\textrm{A}^{\textrm{E}}|k(z)|dz\right)\right)},\nonumber\\
    &=&\frac{T_{sd}}{T_{s, k}},\nonumber\\
    &=&\exp{\left(-\frac{8l}{3\sqrt{3}a\gamma}\left(E_g(1-\epsilon_\kappa) +\phi_{B,n}\right)\right)}, 
\end{eqnarray}
where 
\begin{equation}
    \epsilon_\kappa\equiv \frac{C_s\sqrt{\tilde{\kappa}}\cos{3\alpha}}{\sqrt{3}a\gamma R}\ll 1.
\end{equation}
For $R=4$\AA, for example, $\epsilon_\kappa=0.15\sqrt{\tilde{\kappa}}\cos{3\alpha}$; but $\sqrt{\tilde{\kappa}}\cos{3\alpha}\leq 1$ ($\tilde{\kappa}=1$ corresponds to the kink's closure), hence $\epsilon_\kappa\ll1$ holds everywhere.


The  current per channel is given by
\begin{equation}\label{eq:I(E)}
    I=\frac{2e}{h}\int T(E)[f_s(E)-f_d(E)]dE
\end{equation}
where $f_s(E), f_d(E)$ are the Fermi distributions at the source/drain, respectively.

Near resonance, the total transmission probability  per channel is given by \cite{datta_book},
\begin{equation}\label{eq:T(E)}
    T(E)=\frac{\Gamma_s\Gamma_d}{(E-E_r)^2+\left(\frac{\Gamma_s+\Gamma_d+\Gamma_\phi}{2}\right)^2}
\end{equation}
where $E_r$ is given by eq. \ref{eq:Er}, and $\Gamma_s$, $\Gamma_d$ are the coupling of the kink state to the source/drain, respectively, and $\Gamma_\phi$ is the electron-phonon coupling. The source/drain coupling is given by
\begin{eqnarray}\label{eq:Gamma}
    \Gamma_s&=&\frac{\hbar v_F}{W}T_{s, k} \nonumber\\
    \Gamma_d&=&\frac{\hbar v_F}{W}T_{k, d} 
\end{eqnarray}
where $v_F$ is the Fermi velocity and $W$ is the width of the kink potential well ($\sim 2R$). With $v_F=10^6$m/s, $R=4$\AA \,(for a (10,0) tube), $l=20$nm, $E_g=0.8$eV and $E_\kappa=0.4$eV, eqs. (\ref{eq:Gamma}) give $\Gamma_{s}=7.5\cdot 10^{-6}$eV and 
$\Gamma_{d}=5\cdot 10^{-10}$eV.

The overall current depends sensitively, in addition to the factors treated here, also on the electron-phonon coupling. The peak current, however, is independent of this coupling  \cite{datta_book}, and is given by,
\begin{equation}\label{eq:Ip}
    I_p=\frac{2e}{\hbar} \frac{\Gamma_s\Gamma_d}{\Gamma_s+\Gamma_d}\approx \frac{ev_F T_{k,d}}{R},
\end{equation}
where $\Gamma_s, \Gamma_d$ are given by eq. (\ref{eq:Gamma}), and in the approximation,  $\Gamma_s\gg\Gamma_d$ was assumed, as in eq. (\ref{eq:tau2}).



\subsubsection{Figure of merit}

A simple  measure of sensitivity to the presence of an ion is the quantity
\begin{equation}\label{eq:tau1}
    \tau\equiv\frac{\Delta E_\kappa}{\Gamma}
\end{equation}
 $\tau$ gives the shift in the bound state energy due to an ion in the center of the kink, divided by its width $\Gamma=\Gamma_s+\Gamma_d$  (eq. \ref{eq:Gamma}). Substituting $\Delta E_\kappa$ from eq. (\ref{eq:Delta_E0}), $T_{s,k}$  (\ref{eq:tunnelings_to_kink}) in $\Gamma_s$, and since $\Gamma_s\gg\Gamma_d$, $\Gamma\approx\Gamma_s$, hence,
\begin{equation}\label{eq:tau2}
    \tau=\frac{16k_ee^2E_gV_\kappa R^2}{3\epsilon_r a^2\gamma^2\hbar v_F}\exp{\left(\frac{8lE_gV_\kappa R}{9a^2\gamma^2}\right)},
\end{equation}
or,

\begin{eqnarray}
    \tau&=&\left(a_e a_g x\right) \exp{\left(\frac{8la_g x}{9R}\right)},\,\,\,\,\,\,\textrm{where,}\\
    x&\equiv& \tilde{\kappa}^{1/2}\cos{3\alpha}, \nonumber\\
    a_e&\equiv&\frac{16k_e e^2}{3\epsilon_r\hbar v_F}\approx 1.16,\nonumber\\  a_g&\equiv&\frac{C_sE_g}{a^2\gamma^2},\nonumber 
\end{eqnarray}
where, if $E_g= 1$eV, for example, $a_g=0.14$. $\tau$ is depicted in fig. (\ref{fig:tau}) for a number of junction lengths and bandgaps as a function of post-buckling bending. Clearly, $\tau$ is largest for long junctions and high bandgaps.  However, being a tunneling junction, this also translates to an exponentially  weaker signal (eq. \ref{eq:Ip}).  Thus, $\tau$ and $I_p$ are the two quantities that need to be traded-off.  


\begin{table*}
\begin{center}
\begin{tabular}{ c  c |   c  c c c c c c c c c c c } 
 \hline
  & $n$ &7&8&10&11&13&14&16&17&19&20   \\ [0.5ex] 
 \hline
 &$E_g$&0.474&0.853&0.913&0.945&0.714&0.733&0.586&0.6&0.497&0.5   \\ [0.5ex] 
 
  &$\chi$ &4.803&4.373&4.243&4.187&4.303&4.293&4.367&4.36&4.411&4.407   \\ [0.5ex] 
 \hline
 &$\phi_m$&&&&&$\phi_{B}=\phi_m-\chi$&&&&&   \\ [0.5ex] 
 \hline
 Pt &5.65&0.847&1.276&1.406&1.462&1.347&1.356&1.283&1.29&1.238&1.243   \\ [0.5ex] 
 Pd &5.12&0.317&0.747&0.876&0.932&0.817&0.827&0.753&0.76&0.709&0.713   \\ [0.5ex] 
   Au &5.1&0.297&0.726&0.856&0.912&0.797&0.806&0.733&0.74&0.688&0.693   \\ [0.5ex]  
  Ag &4.26&-0.543&-0.113&0.016&0.072&-0.043&-0.0335&-0.107&-0.1&-0.151&-0.147   \\ [0.5ex]   
  Ga &4.2&-0.603&-0.173&-0.043&0.013&-0.013&-0.093&-0.167&-0.16&-0.211&-0.207   \\ [0.5ex] 
 Mg &3.66&-1.143&-0.713&-0.583&-0.527&-0.643&-0.633&-0.707&-0.7&-0.751&-0.747   \\ [0.5ex]   
  Li &2.9&-1.903&-1.473&-1.343&-1.287&-1.403&1.393&1.467&-1.46&-1.511&-1.507   \\ [0.5ex]   
  K &2.3&-2.503&-2.073&-1.943&-1.887&-2.003&-1.993&-2.067&-2.06&-2.111&-2.107   \\ [0.5ex]    
 \hline
\end{tabular}
\end{center}
\caption{A list of ideal Schottky barriers between various metals and zigzag tubes $(n,0)$. Energy gaps and electron affinity $\chi$ are calculated according to \cite{my_paper_fundamental} while work function ($\phi_m$) are taken from \cite{michaelson}.} 
\label{table:Eg}
\end{table*}

\subsubsection{Conclusion}
We proposed here a solid-state nanopore device based on buckled semiconducting CNT, and a method of fabricating it at scale.
An open kink is an ovalized constriction -- in-effect, a nanopore within a nanotube; in addition, open kinks in semiconducting tubes have a localized bound state within the bandgap. Ions translocating the kink will electrostatically perturb this state. The response of the current to this perturbation is the signal; it can be maximized under certain conditions explored here, in particular, biasing an asymmetric tunnel junction to the negative differential resistance regime,  near the  resonance point.
 
Compared with existing solid-state nanopore devices,  this has two principle novelties: first, it provides
a nano-scale constriction (effectively a nanopore within a nanotube) whose cross-section is smooth and adjustable. And second, the modulated current is not ionic but  electronic.

\appendix
\section{Schottky barriers for zigzag tube}

\begin{figure}[]  
\begin{center}
\includegraphics[width=0.55\textwidth]{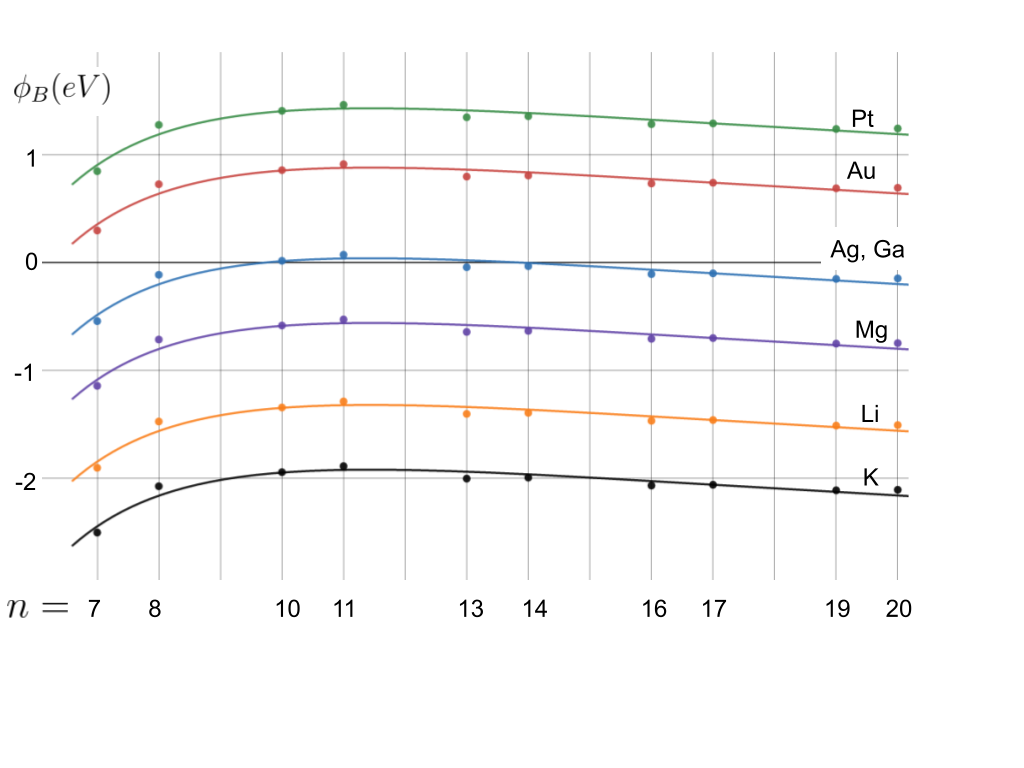}\caption{Ideal Schottky barrier heights $\phi_B$ between semiconducting zigzag tubes and various metals (according to the data in table \ref{table:Eg}). Every number $n$ represents a $(n,0)$ semiconducting  tube. 
} \label{fig:shottky_compare} 
\end{center}
\end{figure}

 This appendix includes  data concerning the \emph{ideal} contacts between a range of semiconducting zigzag tubes and a number of selected metals. It is listed in table \ref{table:Eg} and depicted for a few selected metals in fig. (\ref{fig:shottky_compare}).

The variation in Schottky barrier height among the tubes is due to two factors: bandgap and Fermi level. 
Now bandgap, which is  $\propto 1/R$ for large $n$, increased the electron affinity $\chi$ , by the same measure.
 For small $n$, in contrast, bandgaps are lower but so is the Fermi level (\cite{my_paper_fundamental} and references therein). Hence the lowering of the Schottky barrier height.


\bibliography{kink_bib}
\end{document}